\documentclass[conference]{IEEEtran}
\IEEEoverridecommandlockouts
\usepackage{cite}
\usepackage{amsmath,amssymb,amsfonts}
\usepackage{algorithmic}
\usepackage{graphicx}
\usepackage{textcomp}
\usepackage{xcolor}
\usepackage{diagbox}
\usepackage{threeparttable}
\usepackage{multirow}
\newcommand{\tabincell}[2]{\begin{tabular}{@{}#1@{}}#2\end{tabular}} 

\ifCLASSOPTIONcompsoc
  \usepackage[caption=false,font=footnotesize,labelfont=sf,textfont=sf]{subfig}
\else
  \usepackage[caption=false,font=footnotesize]{subfig}
\fi

\def\BibTeX{{\rm B\kern-.05em{\sc i\kern-.025em b}\kern-.08em
    T\kern-.1667em\lower.7ex\hbox{E}\kern-.125emX}}
\begin{document}

\title{The Model Inversion Eavesdropping Attack in Semantic Communication Systems}

\author{\IEEEauthorblockN{Yuhao Chen, Qianqian Yang$^{\dag}$, Zhiguo Shi, Jiming Chen}
\IEEEauthorblockA{{
% The State Key Laboratory of Industrial Control Technology
}
{College of Control Science and Engineering, Zhejiang University, Hangzhou 310007, China}\\
{College of Information Science and Electronic Engineering, Zhejiang University, Hangzhou 310007, China}\\
{The State Key Laboratory of Industrial Control Technology, Hangzhou 310007, China}\\
% Hangzhou, China \\
\{csechenyh, qianqianyang20$^{\dag}$, shizg, cjm\}@zju.edu.cn
  }
}

\maketitle

\begin{abstract}
In recent years, semantic communication has been a popular research topic for its superiority in communication efficiency. As semantic communication relies on deep learning to extract meaning from raw messages, it is vulnerable to attacks targeting deep learning models. In this paper, we introduce the model inversion eavesdropping attack (MIEA) to reveal the risk of privacy leaks in the semantic communication system. In MIEA, the attacker first eavesdrops the signal being transmitted by the semantic communication system and then performs model inversion attack to reconstruct the raw message, where both the white-box and black-box settings are considered. Evaluation results show that MIEA can successfully reconstruct the raw message with good quality under different channel conditions. We then propose a defense method based on random permutation and substitution to defend against MIEA in order to achieve secure semantic communication. Our experimental results demonstrate the effectiveness of the proposed defense method in preventing MIEA.
\end{abstract}

%\begin{IEEEkeywords}
%model inversion attack, security, semantic communications, deep learning
%\end{IEEEkeywords}

\section{Introduction}
Recently, semantic communication has been widely believed to be one of the core technologies for the sixth generation (6G) of wireless networks because of its high communication efficiency\cite{qin2021semantic}. Compared with the current research on communication which focuses on transmitting mapped bit sequences of the raw message\cite{he2021efficient, ji2021dynamic, xing2022energy}, semantic communication systems transmit compacted semantic features. Existing literature in semantic communication mainly exploits the deep learning (DL) techniques to extract the semantic features from the raw message. For instance, Han \emph{et al}. \cite{han2022semantic} proposed to extract the text-related features from the speech signal as the semantic features and remove the redundant content. On the receiver's side, the semantic features can be reconstructed by a deep learning model into the original message or directly applied for downstream tasks such as image classification and speech recognition.

Although many works have been proposed for semantic communication considering different aspects, few studies have taken into account the security problems\cite{tung2022deep, sagduyu2022semantic, du2022rethinking}. Tung \emph{et al}.\cite{tung2022deep} proposed to encrypt the transmitted signal in semantic communication, but the encryption algorithm incurs a large computation overhead. Security is crucial in semantic communication for two main reasons. Firstly, semantic communication is more prone to privacy leakage compared to traditional communication. In traditional communication systems, the bit sequences being transmitted contain redundant bits to ensure reliable transmission, which can be used to provide a certain level of privacy protection. However, the semantic communication systems transmit compact and more semantic-related symbols which may reveal more private information. Secondly, deep-learning-based semantic communication may be vulnerable to attacks targeting DL models. Extensive studies have been conducted on attacks on the DL model, a review of which can be referred to \cite{liu2020privacy}. If the semantic features being transmitted are eavesdropped by a malicious attacker, the attacker can reconstruct the raw message by utilizing the DL-based attack techniques. The attacker can also add perturbation to the transmitted data, causing the semantic communication system to make incorrect decisions on downstream tasks. For example, Sagduyu \emph{et al}. \cite{sagduyu2022semantic} proposed a multi-domain evasion attack to cause the semantic communication system to make incorrect classifications, which is achieved by introducing noises to input images or the semantic features. Du \emph{et al}.\cite{du2022rethinking} proposed a semantic data poisoning attack, which causes the receiver to receive irrelevant messages from the transmitter. For example, the receiver wants to receive an image with a pear but gets an image with an apple instead. This attack is performed by minimizing the difference between the semantic features of the targeted message and the irrelevant message.
%Therefore, it is essential to consider the security issues in semantic communication.

% Firstly, semantic communication should transmit the semantic meaning over the wireless network, leaving the transmitted data vulnerable to eavesdropping or interference attacks by malicious third parties. In the eavesdropping attack, the attacker intercepts the wireless data being transmitted to steal the sensitive information, while in the interference attack, the attacker disrupts the communication by intentionally transmitting signals on the same frequency band as the wireless signals being transmitted. 

In this paper, we consider the security issue in semantic communication systems and introduce the model inversion eavesdropping attack (MIEA) for semantic communication, where an attacker eavesdrops the transmitted symbols and attempts to reconstruct the original message from them by inverting the DL model used at the transmitter. We perform MIEA under both the white-box and the black-box settings. The attacker has knowledge of the DL model in the white-box setting while not in the black-box setting. To defend against MIEA, we also propose a defense method based on random permutation and substitution. Evaluations demonstrate that the MIEA attack works under different channel conditions, i.e., different values of the signal-to-noise ratio (SNR), which reveals the risk of privacy leaks in semantic transmission. Numerical results also validate the effectiveness of our proposed defense method.

% To reconstruct the raw message, we follow the idea of model inversion attack (MIA) presented in \cite{he2020attacking}, which was initially introduced for edge–cloud collaborative inference. The difference between the edge–cloud collaborative inference and semantic communication is that the channel condition is considered in semantic communication, which may affect the quality of the signal eavesdropped by the attacker. Moreover, we consider the transmission task instead of downstream tasks in this paper, where the current defense methods against MIA are not applicable.

% \begin{itemize}
% 	\item We propose the MIEA, which is a combination of the model inversion attack and the eavesdropping attack, to reconstruct the raw images by two types of attack under different channel conditions.
% 	\item We propose a novel defense method based on random permutation and substitution of the semantic meaning to effectively defend against the MIEA.
% 	\item Evaluations demonstrate the feasibility of MIEA and reveal the risk of privacy leaks in semantic meaning. We also conduct comprehensive experiments to show the effectiveness of our proposed defense method.
% \end{itemize}

This paper is organized as follows: Section \ref{sec fundamentals} introduces the basic ideas of semantic communications. In section \ref{sec miea}, we present the proposed MIEA under both the white-box and black-box setting, and propose our defense method. In section \ref{sec eval}, we evaluate the effectiveness of the proposed MIEA and the proposed defense method. Section \ref{sec conclusion} concludes our work. 

\section{Fundamentals}
\label{sec fundamentals}
\begin{figure}[htbp]
\centering
   \includegraphics[width=3.45in]{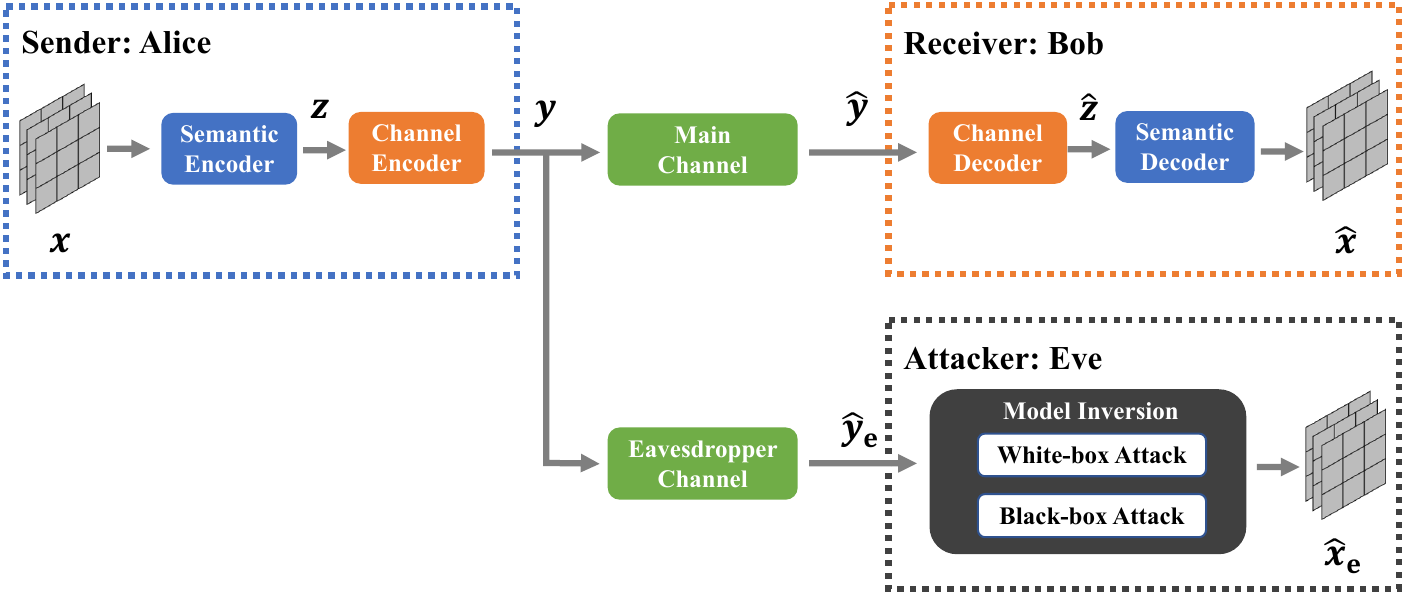}
\caption{Illustration of the semantic communication with MIEA.}
\label{system model}
\end{figure}

In this section, we provide the fundamentals of semantic communication and the eavesdropping performed by the attacker. We consider a semantic communication system which transmits images over wireless channels. As shown in Fig.~\ref{system model}, the transmitter of the semantic communication system consists of a semantic encoder and a channel encoder. The semantic encoder extracts the semantic features $\boldsymbol{z}$ from the raw image $\boldsymbol{x}$, while the channel encoder maps $\boldsymbol{z}$ into the transmitted features $\boldsymbol{y}_\mathrm{f} \in \mathbb{R}^{h \times w \times c}$, where $h, w, c$ denote the height, the width and the channel of the transmitted features respectively. Before transmission, $\boldsymbol{y}_\mathrm{f}$ is reshaped into the transmitted symbols $\boldsymbol{y} \in \mathbb{R}^{N \times 2}$, where $N = \frac{h \times w \times c}{2}$ and the two channels are the real parts and imaginary parts of the signal to be transmitted, respectively. $\boldsymbol{y}$ is then transmitted over a wireless channel, which we denote as the main channel to distinguish from the channel used by the attacker. The received signal $\hat{\boldsymbol{y}}$ at the receiver side can be characterized by 

\begin{equation}
	\hat{\boldsymbol{y}} = \boldsymbol{H}_\mathrm{m}\boldsymbol{y} + \boldsymbol{n}_\mathrm{m},
\end{equation}
where $\boldsymbol{H}_\mathrm{m}$ is a matrix which reflects the main channel effect such as multi-path propagation, fading and interference, while $\boldsymbol{n}_\mathrm{m}$ is a zero-mean additive white Gaussian noise. The receiver of the semantic communication system consists of a channel decoder and a semantic decoder. The receiver first reshapes $\hat{\boldsymbol{y}}$ back to the transmitted features $\hat{\boldsymbol{y}}_\mathrm{f}$. Then the channel decoder maps $\hat{\boldsymbol{y}}_\mathrm{f}$ back to the semantic features $\hat{\boldsymbol{z}}$. The semantic decoder then reconstructs the image $\hat{\boldsymbol{x}}$ from $\hat{\boldsymbol{z}}$. We jointly train the semantic encoder, channel encoder, semantic decoder and channel decoder using the following loss function: 

\begin{equation}
\label{ae loss}
	\mathcal{L} = \frac{1}{N}\sum\limits_{i=1}^N \Vert \boldsymbol{x} - \hat{\boldsymbol{x}} \Vert^2 + \lambda T(\hat{\boldsymbol{x}}),
\end{equation}
where $N$ is the number of the training data batch and 

\begin{equation}
T(\hat{\boldsymbol{x}}) = \sum\limits_{i, j}(|\hat{\boldsymbol{x}}_{i+1,j} - \hat{\boldsymbol{x}}_{i,j}|^2 + |\hat{\boldsymbol{x}}_{i,j+1} - \hat{\boldsymbol{x}}_{i,j}|^2)^{\beta/2}. 
\label{tv loss}	
\end{equation}
\noindent The first term in \eqref{ae loss} computes the mean square error (MSE) between $\boldsymbol{x}$ and $\hat{\boldsymbol{x}}$. The second term $T(\hat{\boldsymbol{x}})$ is the total variation \cite{rudin1992nonlinear} that measures the smoothness of the reconstructed image $\hat{\boldsymbol{x}}$, where $\hat{\boldsymbol{x}}_{i, j}$ denotes the pixel value at the position $(i,j)$ and $\beta$ controls the smoothness of the image, with larger $\beta$ being more piecewise-smooth. The hyper-parameter $\lambda$ balances the two terms. In our work, we choose $\beta = 1$ and $\lambda = 1$.

Next, we introduce how an attacker eavesdrops the transmitted signal under the semantic communication system. We follow the naming convention in the security research, with Alice, Bob and Eve representing the sender, receiver and attacker respectively. Suppose Alice wants to send an image to Bob. As shown in the lower part of Fig.~\ref{system model}, since the transmitted symbols $\boldsymbol{y}$ is transmitted over the wireless channel, it can easily be captured by any unauthorized receiver. Assume that there exists an attacker Eve who intercepts $\boldsymbol{y}$ and attempts to reconstruct the raw image from it. The wireless channel between Alice and Eve is referred to as the eavesdropper channel\cite{mukherjee2014principles}. The received signal at Eve is given by 

\begin{equation}
	\hat{\boldsymbol{y}}_\mathrm{e} = \boldsymbol{H}_\mathrm{e}\boldsymbol{y} + \boldsymbol{n}_\mathrm{e},
\end{equation}
Similarly, $\boldsymbol{H}_\mathrm{e}(\boldsymbol{\cdot})$ represents the eavesdropper channel matrix and $\boldsymbol{n}_\mathrm{e}$ is a zero-mean additive white Gaussian noise. After eavesdropping $\hat{\boldsymbol{y}}_\mathrm{e}$, Eve is able to reconstruct the image, denoted as $\hat{\boldsymbol{x}}_\mathrm{e}$, which will be detailed section \ref{sec miea}. Note that to avoid confusion, we use the \textbf{received image} to denote $\hat{\boldsymbol{x}}$ received by Bob and the \textbf{eavesdropped image} to denote $\hat{\boldsymbol{x}}_{\mathrm{e}}$ eavesdropped by Eve.

% In semantic communication, autoencoder (AE) is one of the most common ways to extract the semantic meaning from the message. An AE consists of an encoder and a decoder, which are deployed at the transmitter side and at the receiver side respectively. Both the encoder and the decoder are comprised of neural networks like the convolutional neural network (CNN) and fully-connected neural network. We follow the naming convention in wireless communication security research, with Alice, Bob and Eve representing the sender, receiver and attacker respectively. In this paper, we consider the wireless transmission task based on semantic communication. As shown in Fig.~\ref{system model}, suppose Alice wants to send an image $\boldsymbol{x} \in \mathbb{R}^{h \times w \times c}$ to Bob, where $w, h, c$ denote the width, the height and the channel of the image respectively. The encoder takes $\boldsymbol{x}$ as the input and outputs the semantic meaning $\boldsymbol{y} \in \mathbb{R}^{h' \times w' \times c'}$. Generally, the total size of $\boldsymbol{y}$ is smaller than $\boldsymbol{x}$, i.e., $h' \times w' \times c' < h \times w \times c$. The semantic meaning $\boldsymbol{y}$ is then transmitted through a noisy wireless channel, which we refer to as the main channel. The noisy semantic meaning $\hat{\boldsymbol{y}}$ received by the receiver via the main channel can be characterized by 

\section{The Proposed MIEA and its Defense}
\label{sec miea}
In this section, we first elaborate the idea of MIEA. To reconstruct $\hat{\boldsymbol{x}}_{\mathrm{e}}$, Eve performs MIA\cite{he2020attacking} using either the white-box attack or the black-box attack, which depends on the knowledge of the semantic encoder and channel encoder that Eve has. Then we propose an effective defense method that defends against both types of attack.

\subsection{White-box Attack} 
\label{sec wb-mia}
In the white-box attack, Eve knows the parameters and structure of the semantic encoder and channel encoder. For example, the semantic communication system is publicly available or available through purchase, such as JPEG. In this case, Eve can directly use the semantic encoder and channel encoder to reconstruct the image. We denote the two encoders as a single function $f(\cdot)$ which maps a given image $\boldsymbol{x}$ to the transmitted symbol $\boldsymbol{y}$, that is, $\boldsymbol{y} = f(\boldsymbol{x})$. The reconstructed image $\hat{\boldsymbol{x}}_\mathrm{e}$ can be obtained by solving the following optimization problem:

\begin{equation}
\hat{\boldsymbol{x}}_\mathrm{e} = \mathop{\arg\min}\limits_{\boldsymbol{x}} \Vert   \hat{\boldsymbol{y}}_\mathrm{e} - f(\boldsymbol{x}) \Vert ^2 + \lambda T(\boldsymbol{x}).
\label{white box atk}	
\end{equation}
The first term in \eqref{white box atk} is the MSE between $\hat{\boldsymbol{y}}_\mathrm{e}$ and $f(\boldsymbol{x})$, while the second term $T(\boldsymbol{x})$ is the total variation defined in \eqref{tv loss} that guarantees the smoothness of $\boldsymbol{x}$. Similar to \eqref{ae loss}, we set $\beta = 1$ and $\lambda = 1$ here. The optimization problem \eqref{white box atk} can be solved by performing the gradient descent, which iteratively updates the input $\boldsymbol{x}$ so that \eqref{white box atk} can be minimized.

\subsection{Black-box Attack}
\label{sec bb-mia}
In the black-box attack, Eve lacks knowledge of the parameters and structures of both encoders. In this case, Eve uses an inverse network of the two encoders, denoted as $f^{-1}(\cdot)$, to inverse $\hat{\boldsymbol{y}}_\mathrm{e}$ back to $\hat{\boldsymbol{x}}_\mathrm{e}$. Specifically, $f^{-1}$ takes $\hat{\boldsymbol{y}}_\mathrm{e}$ as input and outputs $\boldsymbol{x}$, i.e., $f^{-1}(\hat{\boldsymbol{y}}_\mathrm{e}) = \boldsymbol{x}$. To train $f^{-1}$, we assume that Eve can feed a batch of samples $\mathbb{X} = \{\boldsymbol{x}_1, \boldsymbol{x}_2, ..., \boldsymbol{x}_m\}$ into the encoder and capture the corresponding transmitted symbols $\mathbb{Y}_\mathrm{e} = \{\hat{\boldsymbol{y}}_{1}, \hat{\boldsymbol{y}}_2, ..., \hat{\boldsymbol{y}}_m\}$, where $m$ is the number of the samples. Eve then trains $f^{-1}$ using $\mathbb{Y}_\mathrm{e}$ as the input and $\mathbb{X}$ as the ground truth output. We use the $l_2$ norm as the loss function and employ stochastic gradient descent to train the inverse network:

\begin{equation}
f^{-1} = \mathop{\arg\min}\limits_{g}\frac{1}{m}\sum\limits_{i=1}^m \Vert g(\hat{\boldsymbol{y}}_i) - \boldsymbol{x}_i \Vert^2 ,
\label{black box atk}	
\end{equation}
where we use $g$ to represent the inverse network being optimized. Once the inverse network is trained, Eve is able to reconstruct the image from any newly eavesdropped signal.

\subsection{Defense Method against MIEA}
To defend against MIA in deep learning, researchers have proposed various techniques such as differential privacy (DP)\cite{he2020attacking} or attacker-aware training \cite{li2022ressfl}. The DP technique involves adding an additive Laplacian noise to the raw image, while the attacker-aware training adds a regularization term to the loss function during training, which maximizes the MSE between the reconstructed image and the raw image. Both methods prevent Eve from reconstructing high-quality images while maintaining the performance of downstream tasks. However, in the transmission task considered in this paper, both Bob and Eve attempt to reconstruct the image from $\hat{\boldsymbol{y}}$ and $\hat{\boldsymbol{y}}_\mathrm{e}$ respectively. Furthermore, we assume that both the main channel and eavesdropper channel are AWGN channel, i.e., $H_\mathrm{m} = H_\mathrm{e}$. Then the difference between $\hat{\boldsymbol{y}}$ and $\hat{\boldsymbol{y}}_\mathrm{e}$ is $\vert \boldsymbol{n}_\mathrm{m} - \boldsymbol{n}_\mathrm{n} \vert$, which is relatively small. Therefore, if Eve fails to reconstruct high-quality images under the defense methods above, Bob will also fail, which contradicts the goal of the transmission task. To prevent Eve from reconstructing images as well as maintaining the image quality received by Bob, an intuitive solution is to encrypt $\boldsymbol{y}$ using common cryptography algorithms, but this will incur a large computation overhead. To reduce the computation overhead, we propose a defense method based on random permutation and substitution of the transmitted features $\hat{\boldsymbol{y}_\mathrm{f}}$, which can simultaneously defend against both types of attack. 

\emph{Random Permutation and Substitution}. We first introduce the random permutation operation. For the transmitted features $\boldsymbol{y}_\mathrm{f}$, we randomly permute the tensor along the first dimension $h$. We define the permutation scheme $P$ as a random permutation of the array $[0,1,...,h-1]$, where each element represents the index (0-indexed) of $\boldsymbol{y}_\mathrm{f}$. After applying $P$, we obtain the permuted transmitted features $\boldsymbol{y}^\mathrm{p}_\mathrm{f}$. 

Then we perform the substitution operation on $\boldsymbol{y}^\mathrm{p}_\mathrm{f}$ by swapping some of the $y_i$ with $y'_i$ from another transmitted features $\boldsymbol{y}'_\mathrm{f}$, where $i$ in $y_i$ and $y'_i$ indicates that substitution is performed in the same position of both transmitted features. Note that we remove the subscript f in $y_i$ to avoid complex notations. If Alice sends several images to Bob, $\boldsymbol{y}'_\mathrm{f}$ can be the transmitted features of the next image. If Alice only sends one image to Bob, $\boldsymbol{y}'_\mathrm{f}$ can be a random-noise tensor, which will also be sent to Bob. Similarly, we define the substitution scheme $S$ as a sub-array of the array $[0,1,...,h-1]$, where every $i$ in $S$ indicates the $y_{\mathrm{f},i}$ that should be substituted. After substitution, $\boldsymbol{y}^\mathrm{p}_\mathrm{f}$ becomes $\boldsymbol{y}^\mathrm{s}_\mathrm{f}$. 

\begin{figure}[htbp]
\centering
   \includegraphics[width=2.8in]{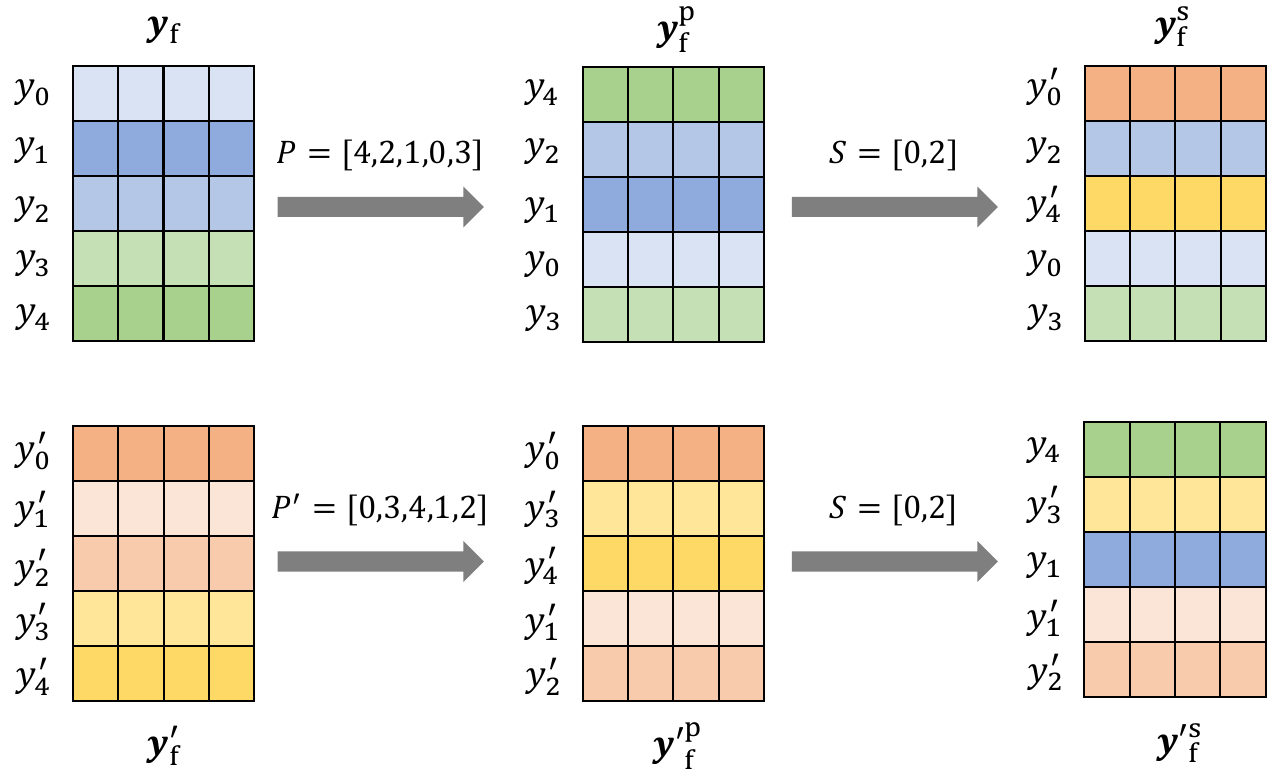}
\caption{An example of the proposed defense method.}
\label{defense eg}
\end{figure}

We give an example to explain the idea of random permutation and substitution. As shown in Fig.~\ref{defense eg}, assume that $h = 5$, then $\boldsymbol{y}_\mathrm{f} = [y_0, y_1, ..., y_4]$, where $y_i \in \mathbb{R}^{1 \times w \times c}$, $i=0,1,...,4$. We also assume that there is another transmitted features $\boldsymbol{y}'_\mathrm{f} = [y'_0, y'_1, ..., y'_4]$  If $P=[4,2,1,0,3]$ for $\boldsymbol{y}_\mathrm{f}$ and $P'=[0,3,4,1,2]$ for $\boldsymbol{y}'_\mathrm{f}$, then $\boldsymbol{y}^\mathrm{p}_\mathrm{f} = [y_4, y_2, y_1, y_0, y_3]$ and ${\boldsymbol{y}'}^\mathrm{p}_\mathrm{f} = [y'_0, y'_3, y'_4, y'_1, y'_2]$. If $\boldsymbol{y}_\mathrm{f}$ and $\boldsymbol{y}'_\mathrm{f}$ share a common $S=[0,2]$, then $\boldsymbol{y}^\mathrm{s}_\mathrm{f} = [y'_0, y_2, y'_4, y_0, y_3]$ and ${\boldsymbol{y}'}^\mathrm{s}_\mathrm{f} = [y_4, y'_3, y_1, y'_1, y'_2]$. Suppose that Bob knows the $P$ and $S$ before transmission. After reshaping $\hat{\boldsymbol{y}}$, Bob will first recover $\hat{\boldsymbol{y}_\mathrm{f}}$ from $\hat{\boldsymbol{y}}^\mathrm{s}_\mathrm{f}$ and then feed $\hat{\boldsymbol{y}_\mathrm{f}}$ into the channel decoder. However, since Eve does not know $P$ and $S$, Eve will try to reconstruct $\hat{\boldsymbol{x}}_\mathrm{e}$ directly from $\hat{\boldsymbol{y}}^\mathrm{s}_\mathrm{f}$, which is shown to be infeasible in \ref{sec eval of defense}. Moreover, since different $P$ and $S$ are used for each transmission, it would be difficult for Eve to determine the correct $P$ and $S$ for each eavesdropped signal. 
%In conclusion, our proposed method protects privacy and maintains the quality of the image received by Bob.

% $\boldsymbol{y}$ is first permuted to $\boldsymbol{y}_\mathrm{p} = [y_5, y_3, y_2, y_1, y_4]$. Then according to $S$, $y_2$ and $y_5$ are substituted with $y'_2$ and $y'_5$ in another $\boldsymbol{y}'$. Finally, the semantic meaning being transmitted is $\boldsymbol{y}_\mathrm{s} = [y'_5, y_3, y'_2, y_1, y_4]$.

\emph{Scheme Selection}. The proposed method is dependent on Bob having knowledge of $P$ and $S$ before transmission. Hence it is necessary for Alice and Bob to share two common sets of schemes, namely the permutation scheme set $\mathbb{P}$ and the substitution set $\mathbb{S}$, which are kept secret from Eve. Both sets comprise multiple schemes that can be employed for permutation and substitution. Before each image transmission, Alice generates a value pair $V = \{p, s\}$, which is used to select the corresponding $P$ and $S$ from $\mathbb{P}$ and $\mathbb{S}$, respectively. $V$ is first encrypted using a secret key $K$ shared between Alice and Bob. Then the encrypted $V$ is transmitted to Bob, which cannot be modified by the main channel. Hence error-free techniques such as error correction and retransmission are utilized to transmit $V$. After receiving the $\hat{\boldsymbol{y}_\mathrm{s}}$ and the encrypted $V$, Bob decrypts $V$ using $K$ and determines $P$ and $S$, from which $\hat{\boldsymbol{y}}$ can be recovered. 

\begin{figure*}[htbp]
\centering
   \includegraphics[width=5.5in]{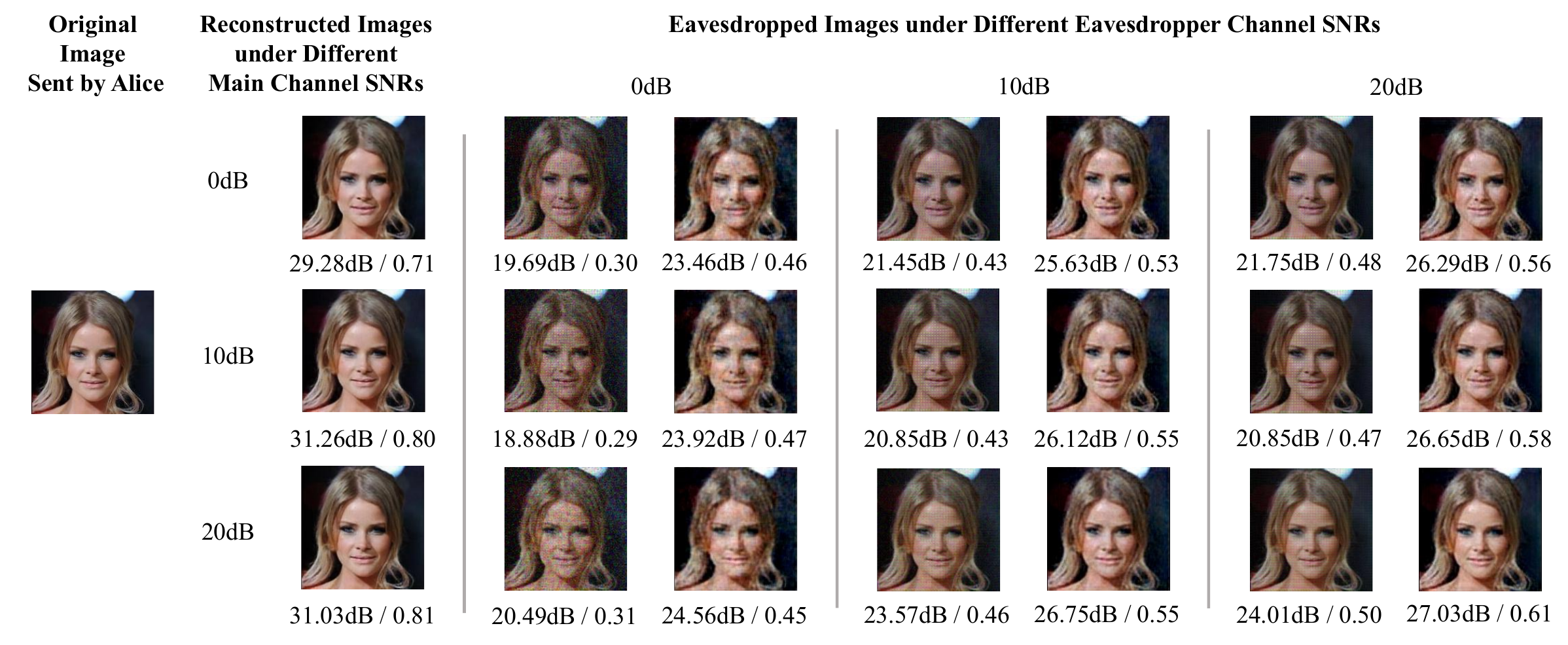}
\caption{Visualization of MIEA for the white-box attack and the black-box attack under different channel conditions. For each channel condition, the eavesdropped image by the white-box attack is displayed on the left and the one obtained by the black-box attack is on the right.}
\label{miea eval fig}
\end{figure*}

\section{Evaluations}
\label{sec eval}
In this section, we present our experiments to evaluate MIEA and the proposed defense method. We first evaluate MIEA's performance for the white-box attack and black-box attack. We then show the effectiveness of the proposed defense method. We use the semantic communication model DeepJSCC\cite{bourtsoulatze2019deep} to transmit images from the CelebA dataset\cite{liu2015faceattributes}, which we crop and resize to $180\times180$ in the evaluation. The semantic encoder and decoder each have four convolutional layers, while the channel encoder and decoder have one convolutional layer. We assume both the main channel and eavesdropper channel to be AWGN channel and denote the channel condition as the combination of the main channel's SNR and the eavesdropper channel's SNR. Although the main channel is not considered in MIEA, we still perform evaluation under different SNRs of the main channel, as we use different DeepJSCC models for each SNR value. For each evaluation, we consider the SNR of both channels to be 0dB, 10dB and 20dB, resulting in nine different channel conditions.

\subsection{Evaluation Setup}
Before evaluating the performance of both attacks, we train the DeepJSCC model on the CelebA dataset using three different SNR values for the main channel (0dB, 10dB, and 20dB), resulting in three distinct DeepJSCC models. As stated in \cite{bourtsoulatze2019deep}, the SNR value determines the standard deviation of $\boldsymbol{n}_\mathrm{m}$ when the transmission power is normalized to 1. We train the CelebA dataset with a batch size of 128, using Adam \cite{kingma2014adam} as the optimizer with a learning rate of $10^{-3}$.

To measure the image quality, we use two metrics, i.e., the structural similarity index measure (SSIM) and the peak signal-to-noise ratio (PSNR)\cite{bourtsoulatze2019deep}, where higher values of SSIM and PSNR indicate better quality.
% of the received images. 
% In the Rayleigh channel, the signal is subject to random amplitude and phase fluctuations due to multipath fading. The Rician channel has a dominant line-of-sight path in addition to the multipath in the Rayleigh channel. 

\subsection{Evaluation of MIEA}
\label{sec eval miea}
We first evaluate MIEA for the two types of attack. For the white-box attack, where Eve reconstructs the image by minimizing \eqref{white box atk}, we employ Adam\cite{kingma2014adam} as the optimizer with a learning rate of $10^{-3}$ and we initialize $\hat{\boldsymbol{x}}_\mathrm{e}$ to an all-zero tensor. For the black-box attack, we use an inverse network $f^{-1}(\cdot)$ consisting of an upsampling layer and two convolution layers. Then we train $f^{-1}(\cdot)$ by solving the optimization problem in \eqref{black box atk}, where we choose the CelebA test dataset as $\mathbb{X}$ and obtain its corresponding transmitted symbols $\mathbb{Y}$. Similarly, we use Adam as the optimizer and set the learning rate to $10^{-3}$.

% \begin{table}[htbp]
%     \caption{The average SSIM and PSNR of the eavesdropped images under different channel conditions}
%     \centering
%     %\resizebox{\linewidth}{!}{
%     \begin{tabular}{c|ccc}
%     \hline
%     &  \multicolumn{3}{c}{\textbf{Main Channel SNR}} \\
%    &\textbf{0dB} &\textbf{10dB} & \textbf{20dB} \\
%     \hline
    
%    \tabincell{c}{\textbf{Reconstructed} \\ \textbf{Images by Bob}} & 30.02dB / 0.70 & 32.28dB / 0.79 & 33.28dB / 0.80 \\
%         \hline

%     \multirow{2}{*}{\textbf{EC SNR 0dB}} & 17.44dB / 0.31 & 16.69dB / 0.30 & 18.58dB / 0.32  \\
%      & 21.65dB / 0.46 & 22.24dB / 0.46 & 22.85dB / 0.46  \\
%     \hline

%     \multirow{2}{*}{\textbf{EC SNR 10dB}} & 18.58dB / 0.40 & 17.80dB / 0.40 & 20.56dB / 0.43 \\
%      & 23.50dB / 0.53 & 23.76dB / 0.54 & 24.33dB / 0.56 \\
%     \hline

%     \multirow{2}{*}{\textbf{EC SNR 20dB}} & 18.74dB / 0.43 & 17.94dB / 0.42 & 20.84dB / 0.46 \\
%      & 23.88dB / 0.57 & 23.98dB / 0.59 & 24.41dB / 0.61 \\
%     \hline
    
%     \end{tabular}
%     \label{avg ssim and psnr for miea}
%     %}
% \end{table}

Fig.~\ref{miea eval fig} shows the performance of MIEA on the DeepJSCC model under different channel conditions, with the SSIM and PSNR given below each image. The first two columns in Fig.~\ref{miea eval fig} are baselines for comparisons with the images eavesdropped by MIEA, where the first column displays the original images $\boldsymbol{x}$ transmitted by Alice, and the second column shows the images received by Bob under different main channel's SNRs. As shown in the first two columns, increasing the SNR improves image quality, as indicated by higher average SSIM and PSNR values. Additionally, higher SNRs reveal more details in the images, such as the female's hair.

The remaining columns in Fig.~\ref{miea eval fig} display the eavesdropped images obtained by MIEA. In the following evaluations in this paper, for each channel condition, we show the eavesdropped image by the white-box attack on the left and the one obtained by the black-box attack on the right. Additionally, Table\ref{avg ssim and psnr for miea} lists the average SSIM and PSNR of the eavesdropped images of individuals selected from the CelebA training set. The first row in the table shows the quality of the images received by Bob, and for each channel condition, the values on the top show the quality of eavesdropped images obtained by the white-box attack, and the values on the bottom show the quality of the eavesdropped images obtained by the black-box attack. We note that we choose CelebA training set for evaluation because the CelebA test set is used for training $f^{-1}(\cdot)$ in the black-box attack. It can be seen from Fig.~\ref{miea eval fig} and Table~\ref{avg ssim and psnr for miea} that the quality of eavesdropped images improves as the SNR of the eavesdropper channel increases for a given SNR of the main channel. Moreover, for a given SNR of the eavesdropper channel, the quality of eavesdropped images is similar under different SNRs of the main channel. It also can be observed that the SSIM and PSNR values in the black-box attack are generally larger than those in the white-box attack. This is because the black-box attack requires training $f^{-1}(\cdot)$ before reconstructing any image from the eavesdropped signal, which needs many samples from $\mathbb{X}$ and $\mathbb{Y}$. In contrast, the white-box attack directly reconstructs the image from the eavesdropped signal without any training in advance. Although the SSIM and the PSNR of the eavesdropped images in both attacks are lower than those of the images received by Bob, the eavesdropped images are visually recognizable and their privacy is compromised, which confirms the effectiveness of MIEA and reveals the risk of privacy leaks in current semantic communication.

\begin{table}[htbp]
\caption{The average SSIM and PSNR of the eavesdropped images under different channel conditions}
\centering
\begin{threeparttable}
    %\resizebox{\linewidth}{!}{
    \begin{tabular}{c|ccc}
    \hline
    &  \multicolumn{3}{c}{\textbf{Main Channel SNR}} \\
   &\textbf{0dB} &\textbf{10dB} & \textbf{20dB} \\
    \hline
    
   \tabincell{c}{\textbf{Reconstructed} \\ \textbf{Images by Bob}} & 30.02dB / 0.70 & 32.28dB / 0.79 & 33.28dB / 0.80 \\
        \hline

    \multirow{2}{*}{\textbf{EC}\tnote{1} \textbf{SNR 0dB}} & 17.44dB / 0.31 & 16.69dB / 0.30 & 18.58dB / 0.32  \\
     & 21.65dB / 0.46 & 22.24dB / 0.46 & 22.85dB / 0.46  \\
    \hline

    \multirow{2}{*}{\textbf{EC SNR 10dB}} & 18.58dB / 0.40 & 17.80dB / 0.40 & 20.56dB / 0.43 \\
     & 23.50dB / 0.53 & 23.76dB / 0.54 & 24.33dB / 0.56 \\
    \hline

    \multirow{2}{*}{\textbf{EC SNR 20dB}} & 18.74dB / 0.43 & 17.94dB / 0.42 & 20.84dB / 0.46 \\
     & 23.88dB / 0.57 & 23.98dB / 0.59 & 24.41dB / 0.61 \\
    \hline
    
    \end{tabular}
    \label{avg ssim and psnr for miea}
    %}
\begin{tablenotes}
    \item[1] EC refers to the eavesdropper channel.
\end{tablenotes}	
\end{threeparttable}
\end{table}

\subsection{Evaluation of the Proposed Defense Method}
\label{sec eval of defense}
Next, we evaluate the proposed defense method by repeating the evaluation of MIEA in section \ref{sec eval miea} with the defense method implemented on $\boldsymbol{y}_\mathrm{f}$. 

%We first demonstrate the effectiveness of our defense method. Then we conduct an ablation study on the permutation-only and substitution-only defense methods.

\begin{figure}[htbp]
\centering
   \includegraphics[width=3.45in]{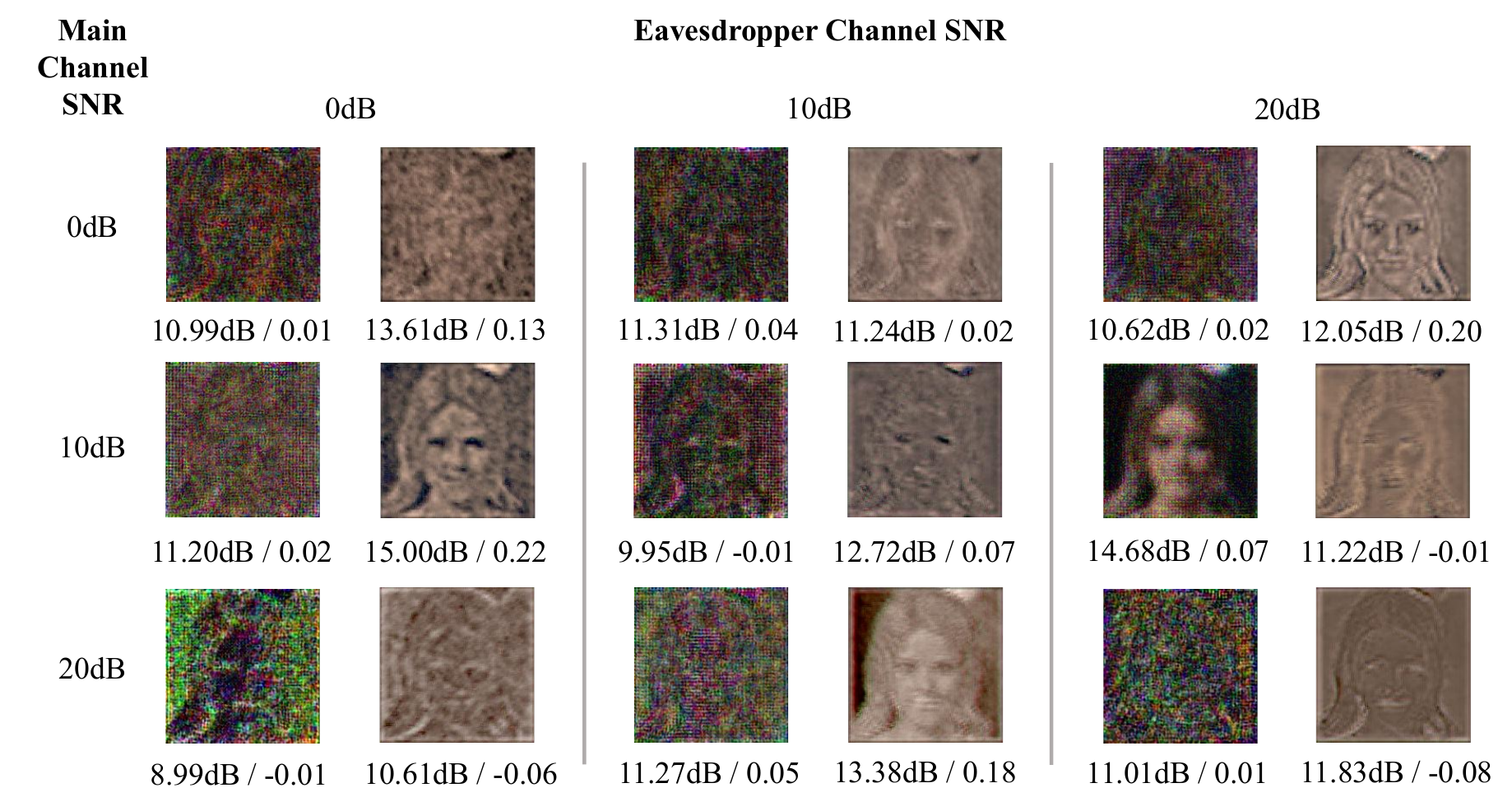}
\caption{Visualization of MIEA for both attacks under different channel conditions after applying the proposed method.}
\label{defense eval}
\end{figure}

Fig.~\ref{defense eval} visualizes the eavesdropped images by the two attack types after applying the proposed defense method, using the same individual as in Fig.~\ref{miea eval fig}. It can be observed that the eavesdropped images are visually unrecognizable, demonstrating the effectiveness of the proposed defense method in preventing Eve from eavesdropping on raw images. We can also see that the contour of the female in the white-box attack is less obvious than that in the black-box attack, suggesting that the defense against the white-box attack is superior to that against the black-box attack. This is because that Eve has no prior knowledge of the defense method when performing the white-box attack, whereas $f^{-1}(\cdot)$ used in the black-box attack has learned some knowledge of the defense method from the training samples. 

% \begin{figure}[htbp]
% \centering
% 	  \subfloat[\label{defense_wb_atk}]{
%       \includegraphics[height=1.8in]{figures/defense_wb_p_s.pdf}}
%     \hfil
%     \subfloat[\label{defense_bb_atk}]{
%       \includegraphics[height=1.8in]{figures/defense_bb_p_s.pdf}}
% \caption{The reconstructed images when the proposed defense method is applied. a) The white-box attack. b) The black-box attack.}
% \label{defense eval}
% \end{figure}

 \begin{table}[htbp]
\caption{The average SSIM and PSNR of the eavesdropped images by MIEA after defense}
\centering
\begin{threeparttable}
    %\resizebox{\linewidth}{!}{
    \begin{tabular}{c|ccc}
    \hline
    \diagbox{EC}{MC\tnote{1}} &\textbf{0dB} &\textbf{10dB} & \textbf{20dB} \\
    \hline
    \multirow{2}{*}{\textbf{0dB}} & 8.03dB / 0.02 & 8.74dB / 0.02 & 6.94dB / 0.00  \\
    & 11.36dB / 0.11 & 11.41dB / 0.07 & 12.51dB / 0.07  \\
    \hline
    
    \multirow{2}{*}{\textbf{10dB}} & 8.55dB / 0.04 & 7.70dB / -0.01 & 9.07dB / 0.05 \\
    & 11.72dB / 0.16 & 11.34dB / 0.11 & 13.22dB / 0.13 \\
    \hline
    
    \multirow{2}{*}{\textbf{20dB}} & 8.02dB / 0.03 & 13.31dB / 0.09 & 8.39dB / 0.02 \\
    & 12.59dB / 0.21 & 11.55dB / 0.10 & 11.54dB / 0.07 \\
    \hline

    \end{tabular}
    \label{avg ssim and psnr for miea after def}
    %}
    \begin{tablenotes}
        \item[1] MC refers to the main channel.
    \end{tablenotes}	
\end{threeparttable}
\end{table}

In addition, we provide the average SSIM and PSNR of the eavesdropped images for both attacks in Table.~\ref{avg ssim and psnr for miea after def}. For a given SNR of the main channel, the SSIM and PSNR do not increase as the SNR of the eavesdropper channel increases because different $P$ and $S$ are used for different transmitted features. The average SSIM and PSNR of the eavesdropped images by the black-box attack are larger than those by the white-box attack, which is consistent with the observation from Fig.~\ref{miea eval fig}. Overall, the average SSIM and PSNR are relatively small, which indicates the effectiveness of the proposed defense method in preventing Eve from obtaining meaningful information from the eavesdropped signal.

%  \begin{table}[htbp]
% \caption{The average SSIM and PSNR of the reconstructed images by the black-box attack after defense}
% \centering
% \begin{threeparttable}
%     %\resizebox{\linewidth}{!}{
%     \begin{tabular}{c|ccc}
%     \hline
%     \diagbox{EC}{MC} &\textbf{0dB} &\textbf{10dB} & \textbf{20dB} \\
%     \hline
%     \textbf{0dB} & 11.36dB / 0.11 & 11.41dB / 0.07 & 12.51dB / 0.07  \\

%     \textbf{10dB} & 11.72dB / 0.16 & 11.34dB / 0.11 & 13.22dB / 0.13 \\

%     \textbf{20dB} & 12.59dB / 0.21 & 11.55dB / 0.10 & 11.54dB / 0.07 \\
%         \hline

%     \end{tabular}
%     \label{avg ssim and psnr for bb atk after def}
%     %}
% \end{threeparttable}
% \end{table}

\begin{figure}[htbp]
\centering
   \includegraphics[width=3.45in]{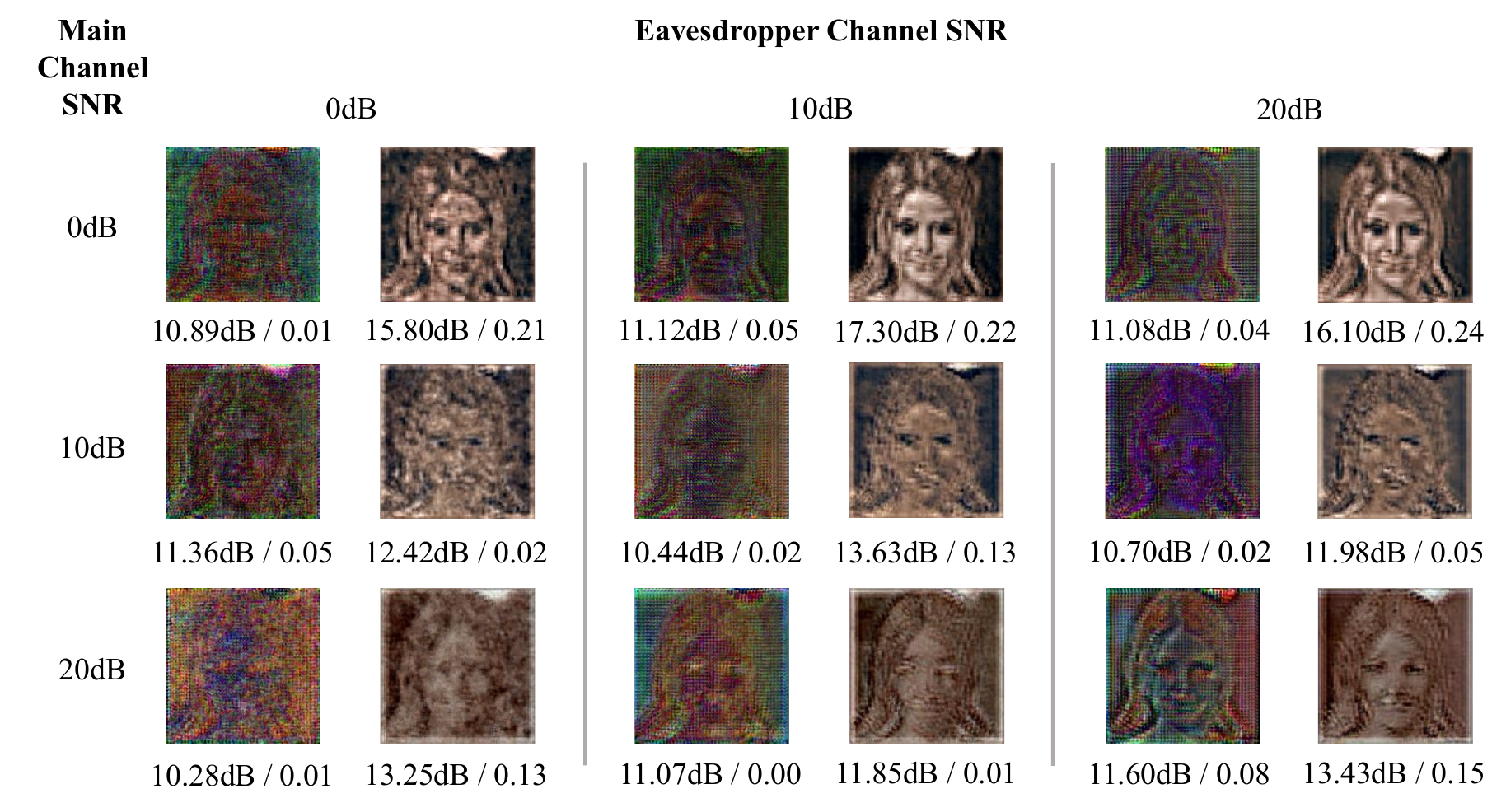}
\caption{Visualization of MIEA for both attacks under different channel conditions after applying only the random permutation.}
\label{defense eval p only}
\end{figure}

 \begin{table}[htbp]
\caption{The average SSIM and PSNR of the eavesdropped images when applying only random permutation}
\centering
\begin{threeparttable}
    %\resizebox{\linewidth}{!}{
    \begin{tabular}{c|ccc}
    \hline
    \diagbox{EC}{MC} &\textbf{0dB} &\textbf{10dB} & \textbf{20dB} \\
    \hline
    \multirow{2}{*}{\textbf{0dB}} & 8.05dB / 0.02 & 8.73dB / 0.06 & 8.50dB / 0.02 \\
    & 14.00dB / 0.22 & 11.43dB / 0.10 & 13.25dB / 0.13  \\
    \hline
    
    \multirow{2}{*}{\textbf{10dB}} & 8.32dB / 0.06 & 8.02dB / 0.03 & 8.87dB / 0.01 \\
    & 14.55dB / 0.26 & 12.97dB / 0.19 & 13.19dB / 0.12 \\
    \hline
    
    \multirow{2}{*}{\textbf{20dB}} & 8.36dB / 0.04 & 8.05dB / 0.02 & 8.77dB / 0.07 \\
    & 14.97dB / 0.28 & 12.83dB / 0.18 & 12.83dB / 0.13 \\
    \hline

    \end{tabular}
    \label{avg ssim and psnr defense p only}
    %}
\end{threeparttable}
\end{table}

% \begin{figure}[htbp]
% \centering
% 	  \subfloat[\label{defense_wb_atk p only}]{
%       \includegraphics[height=1.8in]{figures/defense_wb_p.pdf}}
%     \hfil
%     \subfloat[\label{defense_bb_atk p only}]{
%       \includegraphics[height=1.8in]{figures/defense_bb_p.pdf}}
% \caption{The reconstructed images when only random permutation is applied. a) The white-box attack. b) The black-box attack.}
% \label{defense eval p only}
% \end{figure}

Next, we conduct an ablation study to further validate our proposed method. Fig.~\ref{defense eval p only} and Table.~\ref{avg ssim and psnr defense p only} demonstrate the eavesdropped images and the average SSIM and PSNR for both attacks by applying only the random permutation. As shown in Fig.~\ref{defense eval p only}, when only the random substitution is applied, the white-box attack can be effectively defended, while the black-box attack can still reconstruct visually recognizable images for some $P$ and $S$. Moreover, the average SSIM and PSNR for the black-box attack in Table.~\ref{avg ssim and psnr defense p only} are larger than those in Table.~\ref{avg ssim and psnr for miea after def}, which demonstrates that only random permutation is insufficient for defending against MIEA.

\begin{figure}[htbp]
\centering
   \includegraphics[width=3.45in]{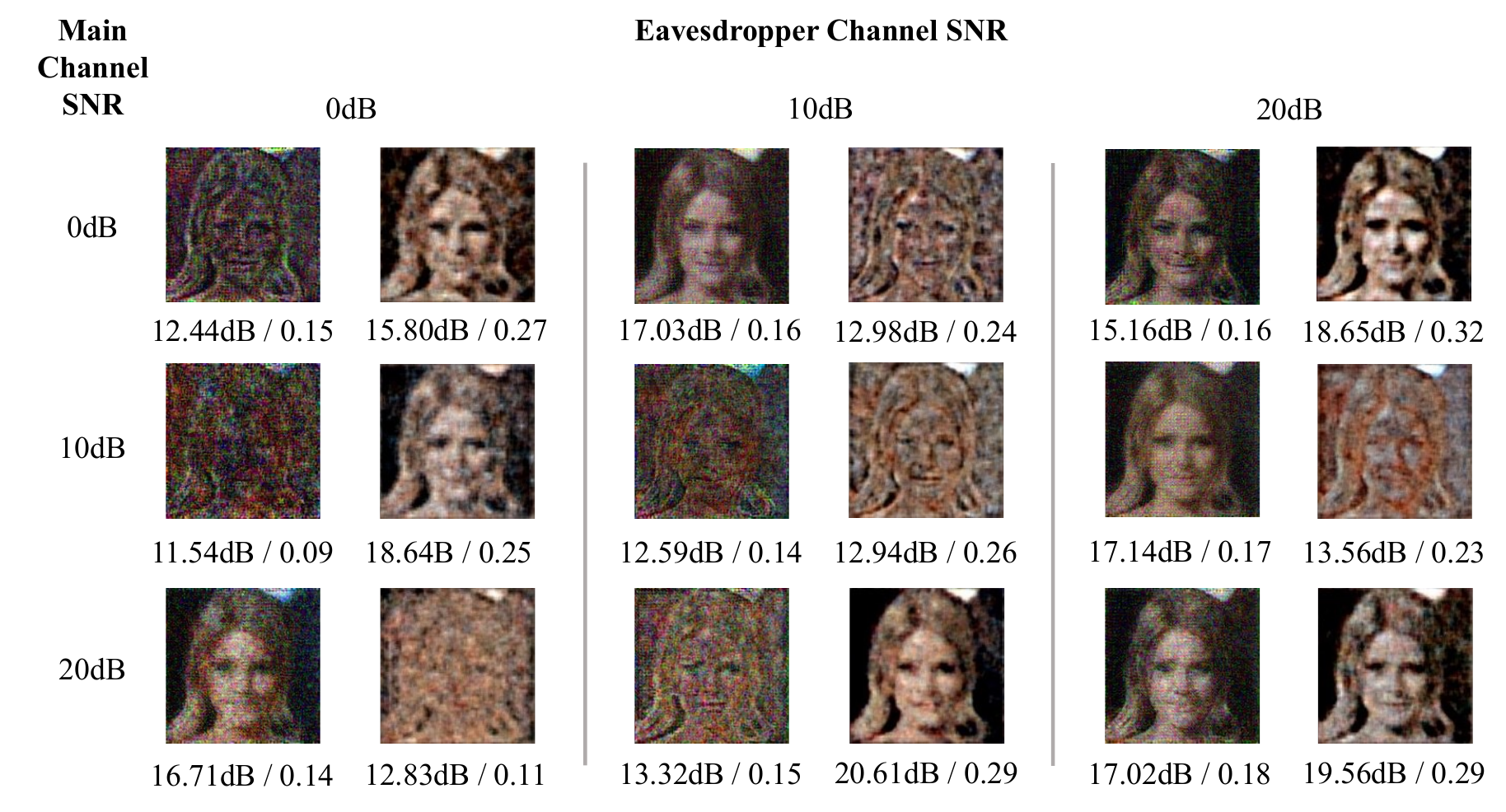}
\caption{Visualization of MIEA for both attacks under different channel conditions after applying only the random substitution.}
\label{defense eval s only}
\end{figure}

Fig.~\ref{defense eval s only} and Table.~\ref{avg ssim and psnr defense s only} show the related results for both attacks by applying only the random substitution. For both attacks, most of the eavesdropped images are visually recognizable, indicating that the attacker can still obtain sensitive information from the transmitted symbols, even though some of the semantic features have been substituted. The average SSIM and PSNR in Table.~\ref{avg ssim and psnr defense s only} are larger than those in Table.~\ref{avg ssim and psnr defense p only}, which means that the random permutation is more effective than random substitution in defending against MIEA. From the ablation study, we can observe that the proposed defense method outperforms both the random-permutation-based and random-substitution-based defense methods, demonstrating that both permutation and substitution are essential for the effectiveness of the proposed defense method.

 \begin{table}[htbp]
\caption{The average SSIM and PSNR of the eavesdropped images when applying only random substitution}
\centering
\begin{threeparttable}
    %\resizebox{\linewidth}{!}{
    \begin{tabular}{c|ccc}
    \hline
    \diagbox{EC}{MC} &\textbf{0dB} &\textbf{10dB} & \textbf{20dB} \\
    \hline
    \multirow{2}{*}{\textbf{0dB}} & 8.99dB / 0.14 & 8.91dB / 0.10 & 15.80dB / 0.16  \\
    & 16.06dB / 0.28 & 15.00dB / 0.25 & 14.70dB / 0.20  \\
    \hline
    
    \multirow{2}{*}{\textbf{10dB}} & 15.62dB / 0.19 & 9.16dB / 0.14 & 10.43dB / 0.16 \\
    & 14.68dB / 0.26 & 14.49dB / 0.26 & 15.80dB / 0.27 \\
    \hline
    
    \multirow{2}{*}{\textbf{20dB}} & 12.68dB / 0.18 & 15.35dB / 0.18 & 16.13dB / 0.21 \\
    & 15.78dB / 0.30 & 14.53dB / 0.27 & 17.29dB / 0.28 \\
    \hline

    \end{tabular}
    \label{avg ssim and psnr defense s only}
    %}
\end{threeparttable}
\end{table}

% \begin{figure}[htbp]
% \centering
% 	  \subfloat[\label{defense_wb_atk s only}]{
%       \includegraphics[height=1.8in]{figures/defense_wb_s.pdf}}
%     \hfil
%     \subfloat[\label{defense_bb_atk s only}]{
%       \includegraphics[height=1.8in]{figures/defense_bb_s.pdf}}
% \caption{The reconstructed images when only random substitution is applied. a) The white-box attack. b) The black-box attack.}
% \label{defense eval s only}
% \end{figure}

\section{Conclusion}
\label{sec conclusion}
In this paper, we propose MIEA to expose privacy risks in semantic communication. MIEA enables an attacker to eavesdrop on the transmitted symbols through an eavesdropper channel and reconstruct the raw message by inverting the DL model employed in the semantic communication system. We consider MIEA under the white-box attack and the black-box attack and propose a novel defense method based on random permutation and substitution to defend against both types of attack. In our evaluation, we first examine MIEA for both attacks under various channel conditions. We then conduct experiments and an ablation study to demonstrate the effectiveness of our proposed defense method.

\bibliographystyle{IEEEtran}
\bibliography{ref.bib}{}

\end{document}